\documentclass[preprint]{revtex4}
\usepackage{epsfig}
\usepackage{color}
\newcommand{\ben}{\begin{eqnarray}}
\newcommand{\een}{\end{eqnarray}}

\newcommand{\bef}{\begin{figure}[h!bt]\centering}
\newcommand{\eef}{\end{figure}}
\newcommand{\bet}{\begin{table}[hbt]\centering}
\newcommand{\eet}{\end{table}}

\begin{document}
\title{Electron doped Ca$_{10}$(Pt$_3$As$_8$)(Fe$_2$As$_2$)$_5$ and Ca$_{10}$(Pt$_4$As$_8$)(Fe$_2$As$_2$)$_5$
- High $T_c$ superconductors with skutterudite intermediary layers}
\author{Ni Ni$^1$, Jared. M. Allred$^1$, Benny. C. Chan$^2$, Robert. J. Cava$^1$ }
\affiliation{
$^1$Department of Chemistry, Princeton University, Princeton, NJ 08544, USA\\
$^2$Department of Chemistry, The College of New Jersey, Ewing, NJ 08628, USA\\
}

\begin{abstract}

It has been argued that the very high transition
temperatures of the highest $T_c$ cuprate superconductors are facilitated by enhanced CuO$_2$ plane coupling through the (Bi,Tl,Hg)-O intermediary layers.
Whether enhanced coupling through intermediary layers can also influence $T_c$ in the iron
arsenide superconductors has never been tested due the lack of appropriate
systems for study. Here we report the crystal structures and properties of
two iron arsenide superconductors, Ca$_{10}$(Pt$_3$As$_8$)(Fe$_2$As$_2$)$_5$ (the ``10-3-8 phase")
and Ca$_{10}$(Pt$_4$As$_8$)(Fe$_2$As$_2$)$_5$ (the ``10-4-8 phase"). Based on -Ca-(Pt$_n$As$_8$)-Ca-Fe$_2$As$_2$-
layer stacking, the most important difference in the structures lies in the structural
and electronic characters of the intermediary platinum arsenide layers. Electron doping
through partial substitution of Pt for Fe in the Fe$_2$As$_2$ layers leads to $T_c$ of 11 K in
the 10-3-8 phase and 25 K in the 10-4-8 phase. Using the chemical concepts of Zintl ion
electron counting and the stability of Pt in the 5$d^8$ configuration we argue that the dramatic difference in $T_c$ arises because the
intermediary layer is semiconducting in the 10-3-8
phase but metallic in the 10-4-8 phase, leading to enhanced interlayer coupling in the 10-4-8 phase. The results suggest that metallic intermediary layers may offer
a new road to higher $T_c$ in iron arsenide superconductors.

\end{abstract}
\pacs{}
\date{\today}
\maketitle
\section{Introduction}

Since the report of superconductivity near 26 K in LaFeAsO$_{0.9}$F$_{0.1}$
in early 2008 \cite{jacs}, several new families of high $T_c$ pnictide superconductors
have been discovered \cite{rotter, 111, 42622b}.
All of these materials are based on layers of edge-sharing FeAs$_4$ and FeSe$_4$ tetrahedra \cite{fese}.
The intermediary layers range from simple alkali or alkaline earth ions \cite{111, rotter}
to more complex but still chemically trivial perovskite layers such as
Sr$_4$V$_2$O$_6$ \cite{42622b}.
Recently, Kudo et al., reported new superconductors with -Ca-(Fe, Pt, As)-Ca-(Fe$_2$As$_2$)-  stacking in the
Ca-Fe-Pt-As system \cite{conference}.
The crystal structures and the character of the intermediary layers were not defined.
This report is intriguing because Fe 122 arsenides, such as BaFe$_2$As$_2$\cite{rotter}, form in the ThCr$_2$Si$_2$ structure type
while Pt 122 arsenides, such as SrPt$_2$As$_2$ \cite{pt122}, form in the CaBe$_2$Ge$_2$ structure type, reflecting a fundamental incompatibility between FeAs and PtAs layers.
Here we identify and characterize the crystal structures and physical properties of two superconductors in the
Ca-Pt-Fe-As system. These materials raise the chemical complexity of the superconducting pnictides to a new level.
The first, Ca$_{10}$(Pt$_3$As$_8$)(Fe$_2$As$_2$)$_5$ (10-3-8), has triclinic symmetry, which is extremely rare among superconductors.
The second, Ca$_{10}$(Pt$_4$As$_8$)(Fe$_2$As$_2$)$_5$ (10-4-8), is a higher symmetry tetragonal phase.
These two superconductors have novel structure types with platinum arsenide intermediary layers based on the skutterudite structure, with formulas
Pt$_3$As$_8$ and Pt$_4$As$_8$ respectively.
As-As dimers are present in these layers and are critical to understanding the electronic state of the superconductors.

For the triclinic 10-3-8 phase, we find that $T_c$ can be tuned from 0 to 11 K
through 6\% to 13\% of Pt substitution on the Fe site in the (Fe$_2$As$_2$)$_5$ layers, which dopes the layers with electrons.
For the tetragonal 10-4-8 phase, 13\% Pt on the Fe sites stabilizes superconductivity at a much higher temperature, 25 K.
Although the ground states of the undoped 10-3-8 and 10-4-8 phases remain unclear, our Pt doping study on
the 10-3-8 phase provides evidence for similarities between this new superconducting family and the 122 as well as 1111 families of arsenide
superconductors.
The Zintl concept and the stability of $d^8$ Pt allow us to conclude that the former phase, with a lower superconducting $T_c$, has a semiconducting
intermediary layer, while the latter, with a 2--3 times higher $T_c$, has a metallic intermediary layer. The difference in metallicity of the
intermediary layers in two arsenide superconductors that are otherwise very similar suggests that this is the factor that determines the difference
 in their $T_c$.
Bearing in mind that doping the FeAs layers
in Fe pnictide superconductors leads to lower $T_c$ than doping the intermediary layers, it may be that $T_c$
can be pushed even higher in these systems through doping on the Ca or Pt$_n$As$_4$ (n=3, 4) layers exclusively.

\section{Methods}
To prepare the crystals, CaAs, FeAs, Fe, Pt and As were mixed in an argon-filled glovebox, pressed into pellets,
and put into alumina crucibles. The crucibles were then sealed in quartz tubes
under 1/3 atmosphere of Ar. For heating temperatures between 700 $^\circ $C and 950 $^\circ $C,
a polycrystalline mixture of one or both superconducting phases together with a considerable amount of PtAs$_2$ was obtained.
For heating temperatures above 1100 $^\circ $C, single crystalline 10-3-8 or 10-4-8 phases intergrown with PtAs$_2$ were obtained.
For the 10-3-8 phase, mm-sized single crystals with no intergrown impurities could be obtained when excess CaAs was added to the mixture,
with the Pt doping concentration manipulated by tuning the Fe to Pt ratio; starting material ratios are summarized in Table I.
The crystal growth tubes were heated to 1100 $^\circ $C - 1180 $^\circ $C, held for one week, furnace-cooled or cooled by 5 $^\circ $C/h to
975 $^\circ $C, and then water quenched. The plate-like 10-3-8 single crystals were then separated by
washing in distilled water. For the 10-4-8 phase, the tubes were
 heated to 1100 $^\circ $C - 1180 $^\circ $C, held for one week, furnace-cooled to
900 $^\circ $C, held for one day, and then water quenched. We were not able to
separate the 10-4-8 phase as sizeable crystals free from PtAs$_2$; growth details for the 10-4-8 phase are also summarized in Table I.
The sample used for single crystal structure refinement of the 10-4-8 phase
was grown from Ru-containing melt. This led to improved crystallinity and
approximately 9\% Ru and 7\% Pt substituted on Fe sites.

Crystal structure determination for both new phases was performed on single crystals at 100 K using a Bruker Apex II single crystal X-ray diffractometer
with graphite-monochromated
Mo K$_{\alpha}$ radiation ($\lambda$ = 0.71073 \AA\ ).
Unit cell refinement and data integration were performed
with Bruker APEX2 software package. Unit cell determination was aided by the CELL NOW program \cite{ShelX}.
The crystal structures were refined using
SHELXL-97 \cite{ShelX} implemented through WinGX \cite{WinGX}.
X-ray powder diffraction patterns were collected on a Bruker D8 Focus diffractometer employing
Cu $K_{\alpha}$ ($\lambda \sim 1.5406$ $\AA$) radiation and a graphite diffracted beam monochromator.
Rietveld refinement was carried out using the FULLPROF program suite \cite{fullprof}. The Bragg peaks were refined using the
Thompson-Cox-Hastings pseudo-Voigt function convoluted with an axial divergence asymmetric peak shape; [001] preferred orientation was included in the
powder refinements.
The elemental analysis was made using energy
dispersive X-ray spectroscopy (EDS) in FEI Quanta 200 FEG Environmental-SEM.

DC magnetization, $M (H)$ and $M (T)$, were measured in a Quantum Design (QD) Magnetic
Properties Measurement System (MPMS) superconducting quantum interface device (SQUID) magnetometer. Heat capacity (relaxation method), AC susceptibility,
resistivity and Hall coefficient measurements were performed
in a QD Physical Properties Measurement System (PPMS). The standard four-probe technique was employed for the resistivity measurements (I=1 $mA$).
A four-wire geometry was used in the Hall coefficient measurements. To remove the magnetoresistive components, the polarity of the
magnetic field ($H \parallel c$) was switched. In both resistivity and Hall effect measurements, the
four thin platinum wires employed were attached to the sample with Epotek H20E silver epoxy.
Seebeck coefficient measurements were performed with a modified MMR Technologies SB100
Seebeck measurement system.

\section{Results}

\subsection{Crystal Structure}
\bef \psfig{file=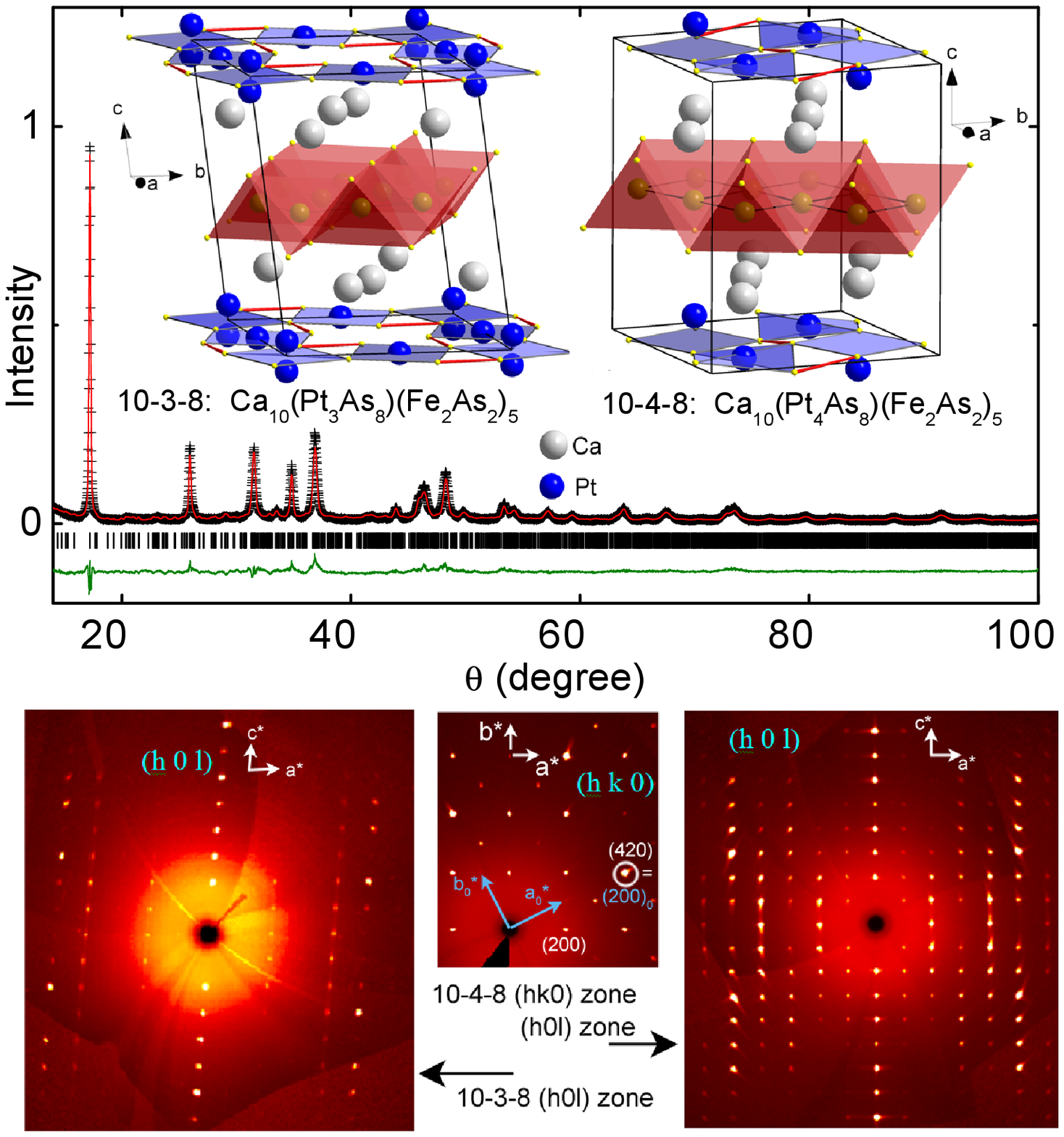,width=3.4in}
\caption{Upper panel: The X-ray powder diffraction pattern of the 10-3-8 phase taken at 300 K. This pattern shows that single phase 10-3-8 can be obtained but it was not employed in the crystal structure determination. Upper black dots, observed pattern; upper red curve,
calculated pattern; tic marks, calculated peak positions using the crystal structure in Table II;
green (lower) curve, difference between observed and calculated pattern. Inset: The crystal structures of the 10-3-8 and 10-4-8 phases.
Lower panels: The single crystal X-ray diffraction patterns for the $(h 0 l)$ zone of the 10-3-8 phase and
the $(h k 0)$ and $(h 0 l)$ zones of the 10-4-8 phase. $a_0^*$, $b_0^*$,
$a^*$ and $b^*$ are described in the text.}
\label{s1}
\eef
The crystal structures of the Ca$_{10}$(Pt$_3$As$_8$)(Fe$_2$As$_2$)$_5$ (10-3-8) and Ca$_{10}$(Pt$_4$As$_8$)(Fe$_2$As$_2$)$_5$ (10-4-8) superconducting
phases,
which have -Ca-(Pt$_n$As$_8$)-Ca-(Fe$_2$As$_2$)- stacking, as well as selected single crystal and powder X-ray diffraction patterns, are shown in
Fig. \ref{s1}.
The detailed crystallographic data determined from the single crystal structure refinements are summarized in Table II and III;
crystals with minimal twinning
were chosen for the structure determinations, and the appropriate twin laws were incorporated into the structure model.
The top panel of Fig. \ref{s1} shows that
the powder X-ray pattern for the 10-3-8 phase can be described well by the determined crystal structure. For both compounds, the strongest reflections reveal a simple tetragonal basal plane subcell with $a_0^*\sim$ 3.91 \AA\
as shown in the $(h k 0)$ single crystal X-ray diffraction pattern.
However, weaker superlattice reflections, which
correspond to a square structural supercell in the real space basal plane, oriented in the (210) direction,
are also observed and lead to $a^* = b^* = a_0^*/\sqrt 5$.
This unusual superlattice condition is mathematically equivalent to that seen in K$_{0.8}$Fe$_{2-\delta}$Se$_2$ \cite{Green},
but arises from a completely different chemical mechanism --- the commensurability condition of the (Pt$_3$As$_8$) or (Pt$_4$As$_8$) skutterudite
layers with the Fe$_2$As$_2$ layers, rather than partial vacancy ordering.
Inspection of the single crystal diffraction patterns in the $c^*$ direction indicates that 10-4-8 phase has a primitive tetragonal cell, while
the 10-3-8 phase shows significant shifting of the stacking between neighboring layers. This layer shift results in a triclinic unit cell,
despite the fact that the basal plane cell is essentially square, i.e., with $a = b$ and $\gamma \approx 90^\circ $.

\bef \psfig{file=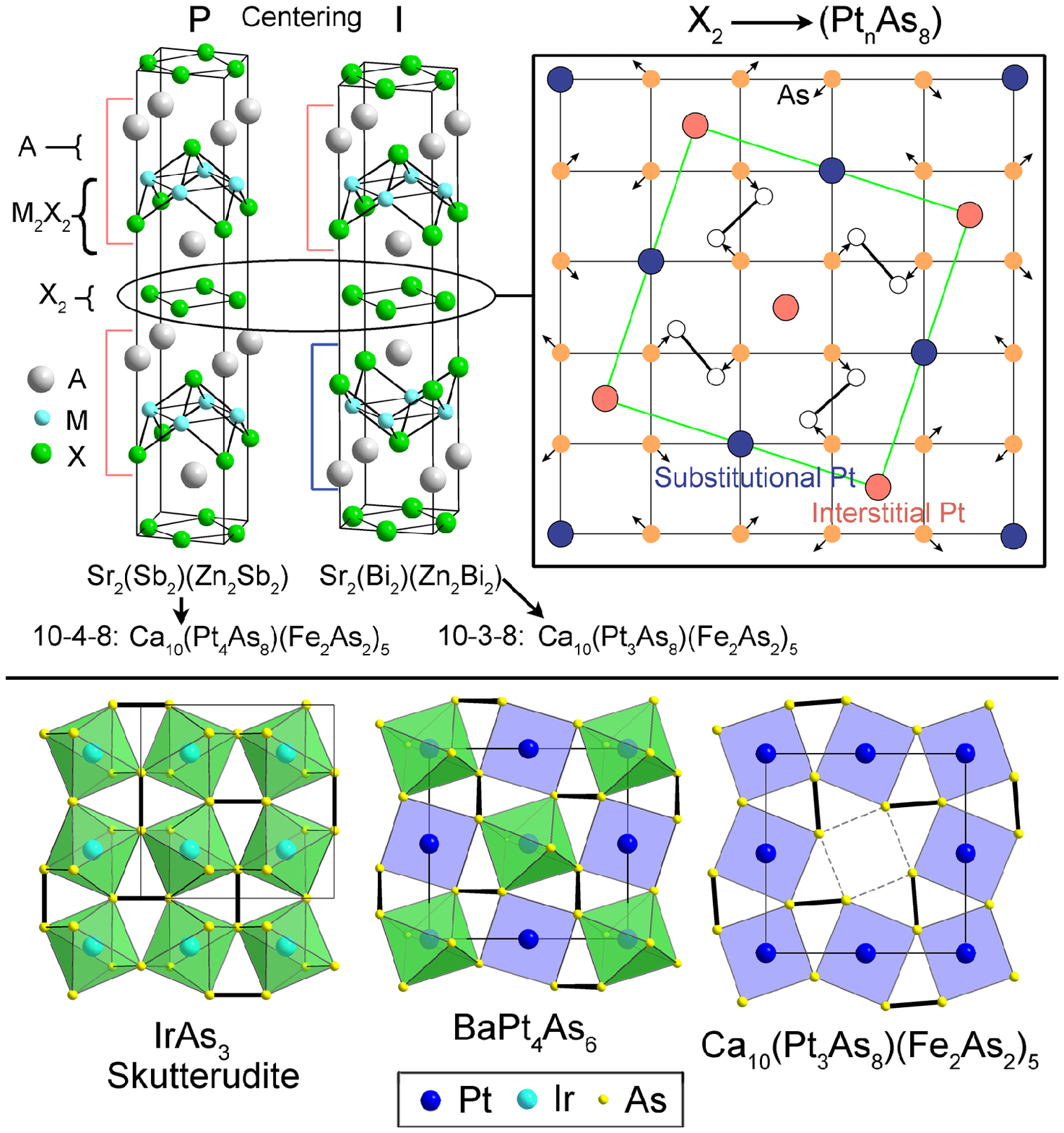,width=3.4in}
\caption{Top left: The crystal structures of SrZnSb$_2$ and SrZnBi$_2$ \cite{hoffman}, which can be considered parent phases of the new Fe-Pt-As superconductors.
Top right: schematic illustration of the transformation from the X$_2$ layer in the parent phases to the
Pt$_n$As$_8$ skutterudite-like layers in the superconductors. Displacements are exaggerated for clarity.
Bottom: Comparison of the crystal structures of skutterudite-structure IrAs$_3$ \cite{iras3}, the platinum arsenide layer in BaPt$_4$As$_6$ \cite{baas}, and the Pt$_3$As$_8$
intermediary layer in the 10-3-8 superconducting phase. As-As bonds are shown bolded.}
\label{s2}
\eef

The 10-3-8 and 10-4-8 phases
are new structure types that can be classified as derivatives of the SrZnBi$_2$ and SrZnSb$_2$ structures respectively \cite{hoffman},
as shown in the top left panel of Fig. \ref{s2}.
As variants of the more common ThCr$_2$Si$_2$ structure,
which accounts for many of the high $T_c$ Fe pnictide superconductors,
the SrZnSb$_2$ and SrZnBi$_2$ structure types have every other M$_2$X$_2$ tetrahedral layer in AM$_2$X$_2$ replaced with a square X$_2$ layer
of the same size, as shown in Fig. \ref{s2}.
The only difference between the SrZnSb$_2$ and SrZnBi$_2$ structures is that one is body
centered while the other is primitive.
This difference in centering is allowed chemically because the X$_2$ layer is not sensitive
to the adjacent A cation arrangement,
resulting in the fact that both staggered and nonstaggered stackings can form around this layer. The difference
in stacking has a profound effect in the superconducting phases,
in which the X$_2$ layer is replaced by a more complex Pt-As layer that is very sensitive to the stacking difference.

The X$_2$ layers are transformed to Pt$_3$As$_8$ and Pt$_4$As$_8$ layers
in Ca$_{10}$(Pt$_3$As$_8$)(Fe$_2$As$_2$)$_5$ and Ca$_{10}$(Pt$_4$As$_8$)(Fe$_2$As$_2$)$_5$ respectively.
This transformation is illustrated schematically in the top left panel in Fig. \ref{s2}.
Starting from the square lattice of As atoms, the substitution
of 1/5 of the As with Pt and the insertion of interstitial Pt leads to strong displacements of the As from their ideal positions.
This occurs so that intralayer As-As dimers are formed and new Pt$_n$As$_8$ (n=3, 4) skutterudite-like layers emerge.
The periodicity of these layers is based on the size of the
the Pt sublattice, which matches the FeAs lattice size with $a = \sqrt 5 \times a_{\rm{Fe_2As_2}}$.
Thus, the Fe$_2$As$_2$ and Pt$_n$As$_8$ layers become commensurate in the (210) Fe$_2$As$_2$ direction (Fig. \ref{s2} upper right).
The resulting Pt$_3$As$_8$ intermediary layer for the 10-3-8 superconductor is illustrated in the right bottom panel of Fig. \ref{s2}.
The layer consists of a square lattice of corner-sharing PtAs$_4$ squares
with a rotation of $\sim 25^\circ$ about an axis perpendicular to the plane, governed by the formation of
intraplanar As-As dimers. This arrangement of atoms is unique in the superconducting iron pnictides, but is fairly
common in platinum group pnictides. A simple example of a compound where such rotations of corner shared MX$_4$ squares is
dictated by the formation of pnictide-pnictide bonds is skutterudite IrAs$_3$ \cite{iras3}, shown in the lower panel of Fig. \ref{s2}.
In this compound the arrangement of As atoms has the same
projected in-plane structure
but all the Ir atoms are octahedrally coordinated rather than having the lower coordination observed for Pt in the 10-3-8 and 10-4-8 phases.
Platinum-based compounds with very similar structures have also been observed,
such as BaPt$_4$As$_6$ \cite{baas}, shown in Fig. \ref{s2}, where 1/2 of the Pt are octahedrally coordinated and the rest are square planes.
In both cases the rotations have an out of plane component, which causes the spacing between Pt atoms to contract.
The As-As bond distance in the dimers is quite similar in these two compounds, 2.47 \AA\ - 2.54 \AA\ in IrAs$_3$ and 2.41 \AA\ - 2.42 \AA\ in BaPt$_4$As$_6$.
These distances are comparable to very similar distances in the new Ca-Fe-Pt-As superconductors
where they are 2.48 \AA\ - 2.49 \AA\ in the 10-3-8 phase and 2.49 \AA\ in the 10-4-8 phase.

\bef \psfig{file=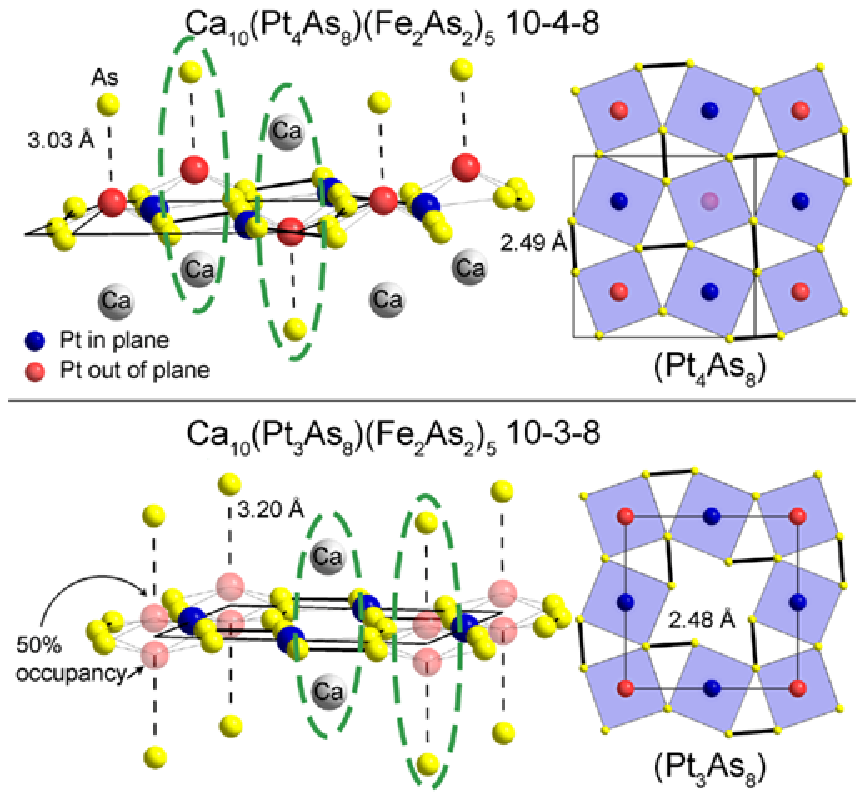,width=3.4in}
\caption{Top: The detailed structure of the Pt$_4$As$_8$ layer in the 10-4-8 superconductor phase Ca$_{10}$(Pt$_4$As$_8$)(Fe$_2$As$_2$)$_5$
Bottom: The detailed structure of the Pt$_3$As$_8$ layer in the 10-3-8 superconductor phase Ca$_{10}$(Pt$_3$As$_8$)(Fe$_2$As$_2$)$_5$.
 The dashed lines show the out-of-plane Pt-As bonding. The dashed ovals show the regions of Pt-Ca positional conflict.}
\label{s3}
\eef
The limited range of the As-As bond lengths present in the new superconductors indicates
that the size of the Pt sublattice is constrained by both the As-As dimer size and the Fe$_2$As$_2$ sublattice size.
The net result of these strong constraints is that only 1/2 of the total number of square sites in the plane are large enough to contain a
Pt atom in a simple square planar coordination with As. This means that Pt
cannot sit exactly in the middle of the remaining squares, but instead must be displaced to a position about 0.5 \AA\ above or below the plane to have favorable Pt-As bond lengths.
This displacement causes a conflict with the Ca ions that are adjacent to the Pt-As layers, resulting in the difference in symmetry
and formula for the primitive and body centered superconducting phases.  This situation is illustrated in Fig. \ref{s3},
which compares the Pt$_3$As$_8$ and
Pt$_4$As$_8$ layers in detail. The Pt atoms shown in blue are present in both Pt$_3$As$_8$ and
Pt$_4$As$_8$ layers and are the ones called "substitutional" Pt atoms in
Fig. \ref{s2}. The Pt atoms shown in red (the ones called "interstitial" Pt atoms in Fig. \ref{s2})
are critical to the Pt stoichiometry of the layers. The differences can be seen by focusing on the parts of the structures that are encircled by
the dashed ovals in the figure.
In the primitive 10-4-8 case (top panel), for each interstitial Pt site, only one side of the plane
is blocked by Ca,
resulting in rooms for a single Pt atom above the plane on one site and below the plane on the other.
In the 10-3-8 case, on the other hand (bottom panel),
one of the potential interstitial Pt sites is blocked on both sides of the plane by the presence of Ca ions, and thus no Pt atoms
can sit on these sites; for the other potential interstitial sites, however, Ca does not interfere, and Pt can occupy a position on either side
of the plane. Pt cannot occupy both sides of the plane at once, however, because then the Pt-Pt separations would be too small; thus the
above-plane and below-plane sites are randomly occupied with a 50\% probability.
We do not have any evidence that there is any long range ordering to lift this disorder.
This difference in the blocking of the interstitial Pt sites, due to the different arrangements of
neighboring Ca, accounts for the difference in formulas of the two new superconductors because
the 10-3-8 structure can only accommodate filling of 1/2 of the interstitial Pt
sites while the 10-4-8 structure allows filling of all of them.

\subsection{Physical Properties}
 In all preparations of the 10-3-8 and 10-4-8 superconductors, we found some fraction of Pt doped on the Fe site in the Fe$_2$As$_2$ layers.
 EDS (Energy dispersive X-ray spectroscopy) measurements were used to determine the Pt doping concentrations on the Fe sites
 through measuring the percentages of elements present in the resistivity sample.
 Structure refinements confirm the Pt doping in the layers and the fact that Fe
  is not found in the Pt-As intermediary layers.
 (To easily compare the physical properties of the new superconductors with the other
 Fe pnictide superconductors, which have much simpler formulas, the units of molar
susceptibility, magnetization and heat capacity presented are normalized to one (Fe$_{1-x}$Pt$_x$)$_2$ per formula unit.)
The EDS measured $x$ values are summarized in Table I.

\bet
\begin{tabular}{c | c | c | c| c|c}
   \hline

  phase & starting ratio & EDS & $x^*$   & $T_c^R$ (K) &$T_c^M$ (K)\\
  ~         & Ca:Fe:Pt:As&  Ca:Fe:Pt:As    & ~  &~\\
\hline
   10-3-8  & 2 : 2 : 0.4 : 4 & 2 : 1.93(3) : 0.70(1) : 3.86(5) &0.06(1)&0 &0\\

   ~ & 2 : 2 : 0.5 : 4 & 2 : 1.89(3) : 0.73(1) : 3.83(5) &0.07(1)&5.9 &4.9\\

   ~ & 4.3 : 2 : 0.7 : 6.3 & 2 : 1.86(2) : 0.77(1) : 3.88(2) &0.09(1)&9.6 &8.2 \\
    ~ & 3 : 2 : 1 : 4 & 2 : 1.79(4) : 0.88(2) : 3.7(1) &0.13(1)&9.9 &10.6\\
\hline
10-4-8 & 2 : 1.5 : 1.5 : 4 & 2 : 1.77(4) : 1.08(1) : 3.77(6) &0.13(1)&22.5 & 24.6\\

     \hline
   \hline
    \end{tabular}
\caption{The starting material ratio in the sample growth, the EDS measured ratio of Ca:Fe:Pt:As and $x$*: the resulting $x$ in
Ca$_{10}$(Pt$_3$As$_8$)((Fe$_{1-x}$Pt$_x$)$_2$)$_5$ for the 10-3-8 phase and Ca$_{10}$(Pt$_4$As$_8$)((Fe$_{1-x}$Pt$_x$)$_2$)$_5$ for the 10-4-8 phase. $T_c^R$ is the $T_c$ obtained from the 50\% criterion from the resistivity data. $T_c^M$ is the $T_c$ obtained from the AC susceptibility data.}
\eet

\bef \psfig{file=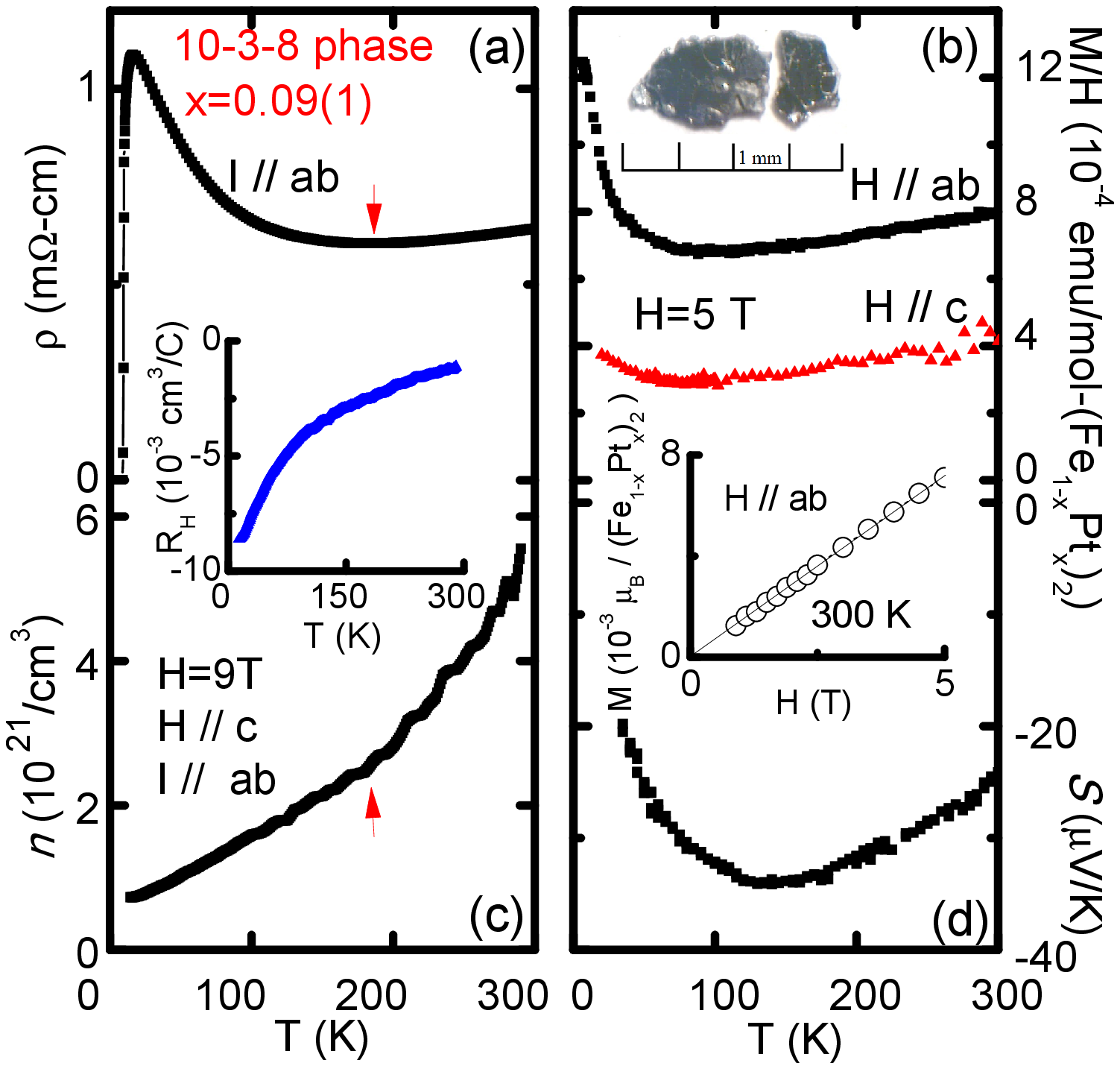,width=3.4in}
\caption{The characterization of the normal state properties for the 10-3-8 phase Ca$_{10}$(Pt$_3$As$_8$)((Fe$_{1-x}$Pt$_x$)$_2$As$_2$)$_5$ with $x=0.09(1)$. (a) Temperature-dependent in-plane electrical resistivity $\rho$.
(b) The temperature dependent magnetic susceptibility, $M(T)/H$, taken at 5 T with $H \parallel ab$ and $H \parallel c$. Inset: the field dependent
magnetization data, $M(H)$, taken at 300 K with $H \parallel ab$. Inset: 10-3-8 single crystals and 1 mm scale.
(c) The estimated temperature dependent carrier density $n$. Inset: the temperature dependent Hall coefficient $R_H$.
(d) The temperature dependent in-plane Seebeck coefficient, $S$.}
\label{NN6591}
\eef
As a representative of the 10-3-8 phase, the picture of the mm-size single crystals and the physical properties of the $x=0.09(1)$
sample are shown in Fig. \ref{NN6591}.
Figure \ref{NN6591} (a) shows the electrical resistivity, $\rho(T)$, from 2 to 300 K. As temperature decreases,
the resistivity slowly decreases from 0.7 m$\Omega$-cm at 300 K to 0.61 m$\Omega$-cm near 190 K, indicated with the red arrow,
and then increases monotonically
to 1.1 m$\Omega$-cm at 12 K, followed
by a sharp decrease to zero at the superconducting transition. The magnitude of the normal state resistivity is similar to that
seen in other Fe pnictide superconductors,
an indication of the ``poor metal" nature of the phase. Figure \ref{NN6591} (b) presents the normal-state magnetic properties.
At 300 K, the magnetization $M$ is linearly proportional
to H, indicating that no ferromagnetic impurities are present in the single crystal sample.
Thus the temperature dependent susceptibility, $\chi(T)$, was measured at 5 T and calculated as $M(T)/H$. Magnetic
anomalies, such as are often associated with structural or magnetic phase transitions in the pnictide superconductors,
were not observed for the 10-3-8 compound, with either H $\parallel$ $ab$ or H $\parallel$ $c$.
The upturn in $\chi$ below 80 K may be attributed to paramagnetic impurities in the sample. From 80 to 300 K, $\chi(T)$
increases approximately linearly with temperature. The ratio of $\chi^{\parallel ab}$ over $\chi^{\parallel c}$ at 300 K is $\sim$ 2.
This value is larger than the ratio of 1.6 observed in CaFe$_2$As$_2$, indicating higher anisotropy in the 10-3-8 superconductor.
The Hall coefficient, $R_H$, shown in the inset of Fig. \ref{NN6591} (c), is negative at all temperatures,
indicating that electron carriers are dominant in this compound.
If a single band model is assumed, then the carrier concentration $n (T)$ can be estimated as $-1/eR_H$. This is plotted as
a function of temperature in Fig. \ref{NN6591} (c).
$n$ decreases from 5.5 $\times 10^{21} \rm {cm}^{-3}$ at 300 K (3.2 times of that of LaFeAsO$_{0.89}$F$_{0.11}$)
to 0.74 $\times 10^{21} \rm {cm}^{-3}$ at 12 K.
A slope change is observed in $n(T)$ at around 190 K, indicated by the red arrow, at the same temperature where the
minimum normal-state resistivity is found (Fig. \ref{NN6591} (a)).
The Seebeck coefficient, shown in Fig. \ref{NN6591} (d), is negative throughout the measured temperature range,
with a room temperature value
of -24.3 $\mu$V/K and minimum value of -34.3 $\mu$V/K near 150 K; this again indicates the dominant role of electrons in the transport.

\bef \psfig{file=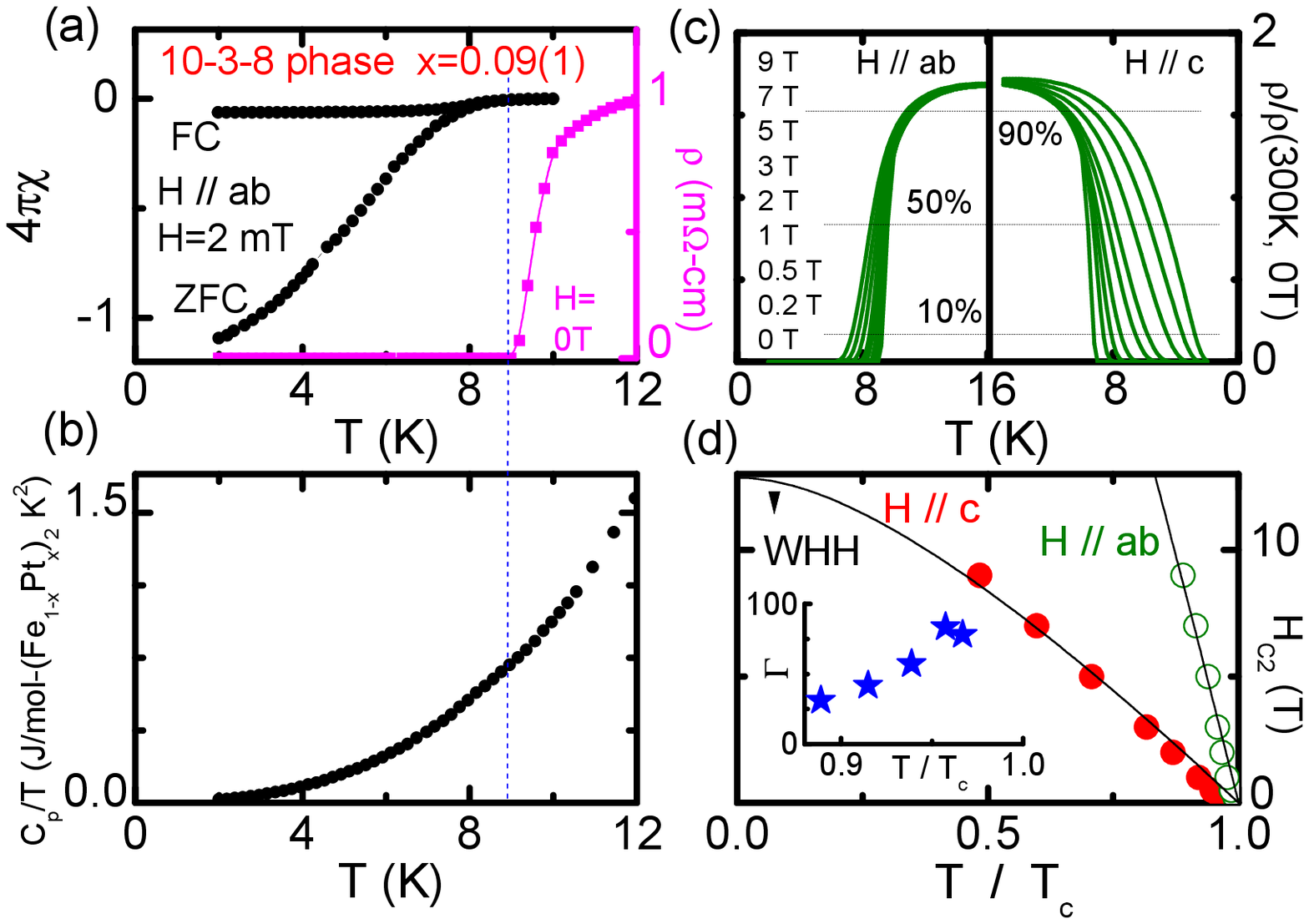,width=4in}
\caption{The characterization of the superconducting properties for the 10-3-8 phase Ca$_{10}$(Pt$_3$As$_8$)((Fe$_{1-x}$Pt$_x$)$_2$As$_2$)$_5$ with $x=0.09(1)$.
(a) Enlargement of the temperature dependent electrical resistivity near $T_c$
and the
field-cooled / zero-field-cooled DC susceptibility $M(T)/H$ with $H \parallel ab$.
(b) $C_p(T)/T$ vs. T. (c) The normalized resistivity, $\rho/\rho(300~ \rm K, 0~ \rm T)$ taken at 9, 7, 5, 3, 2,
1, 0.5, 0.2 and 0 T with $H \parallel ab$ and
$H \parallel c$. The 10\%, 50\% and 90\% criteria are shown.
(d) The inferred anisotropic $H_{c2} (T)$ for $H \parallel ab$ and
$H \parallel c$ using the 50\% criterion.
The solid line is the single band WHH fit \cite{WHH}. Inset:
the anisotropic parameter $\Gamma (T)=(H_{c2}^{//ab}/H_{c2}^{//c})^2$.}
\label{NN659sc}
\eef

Figure \ref{NN659sc} presents the physical properties of the superconducting state for the triclinic 10-3-8 phase.
The zero-field-cooled (ZFC) and field-cooled (FC) DC susceptibility measurements were performed at 2 mT with $H\parallel ab$
so that the demagnetization effect could be minimized.
The diamagnetic signal observed below 9 K in both ZFC and FC measurements confirms
the bulk superconductivity in this compound, and is consistent with the resistivity measurements.
The shielding fraction estimated from the
ZFC data is around 120\%, similar to what is observed in transition metal doped BaFe$_2$As$_2$ \cite{nico}.
The Meissner fraction inferred from the FC data is only about 6\% due to the flux pinning,
which is the usual case in the Fe pnictide superconductors \cite{nico}.
The temperature dependent $C_p(T)/T$ data is presented in Fig. \ref{NN6591} (b).
A subtle heat capacity jump associated with the superconducting transition is observed near 9 K, confirming the bulk superconductivity.
$\rho (T)/\rho (300~ \rm K, 0 ~\rm T)$, measured at 9, 7, 5, 3, 2, 1, 0.5, 0.2, 0 T with $H \parallel ab$ and $H \parallel c$ is presented in Fig. \ref{NN659sc} (c).
With applied field, $T_c$ is suppressed to lower temperatures and the resistive transition broadens, indicating the presence of
strong thermal fluctuation of the vortices.
This is different from what is observed in the Ba(Fe$_{1-x}$Co$_x$)$_2$As$_2$ superconductors, where
no broadening was observed \cite{nico}, but is reminiscent of that in RFeAs(O$_{1-x}$F$_x$) and cuprates \cite{naturehc2}.
To determine $T_c$ at each field,
90\%, 50\% and 10\% of the normal-state resistance at 16 K are used as the criteria.
At 0 T, $T_c^{90\%}$ = 11.31 K, $T_c^{50\%}$ = 9.64 K and $T_c^{10\%}$ = 9.22 K.
For all three criteria, the $H_{c2}^{//ab}$ curves show roughly linear behavior.
The $H_{c2}^{//c}$ curve, however, changes from a concave shape for the 90\% criterion to a convex shape for the 10\% criterion.
As a compromise, we focus on the H$_{c2}$ values inferred from the 50\% criterion, shown in Fig. \ref{NN659sc} (d).
With 9 T applied field, the $T_c$ was suppressed to 0.9$T_{c0}$ with $H \parallel ab$ and 0.5$T_{c0}$ with $H \parallel c$.
The single band WHH theory, without taking into account the effects of spin paramagnetism and spin-orbit scattering
is used to fit the $H_{c2}$ curves \cite{WHH}. This fit is shown as
the solid curve in the panel. The resulting $H_{c2}^{//ab}(0)$ = 55 T, and $H_{c2}^{//c}(0)$ = 13 T. The anisotropy parameter
$\Gamma=m_c^*/m_{ab}^*=(H_{c2}^{//ab}/H_{c2}^{//c})^2$ decreases from 100 near $T_{c0}$ to 25 at 0.9$T_{c0}$,
as shown in the inset of Fig. \ref{NN659sc} (d).
In accordance with the Ginzburg-Landau theory, $H_{c2}^{//c}=\phi_0/2\pi \xi_{ab}^2$ and $H_{c2}^{//ab}=\phi_0/2\pi \xi_{ab}\xi_{c}$, the
coherence lengths are estimated to be $\xi_{ab}(0)$= 50 \AA\ and $\xi_{c}(0)$= 12 \AA\ .

\bef \psfig{file=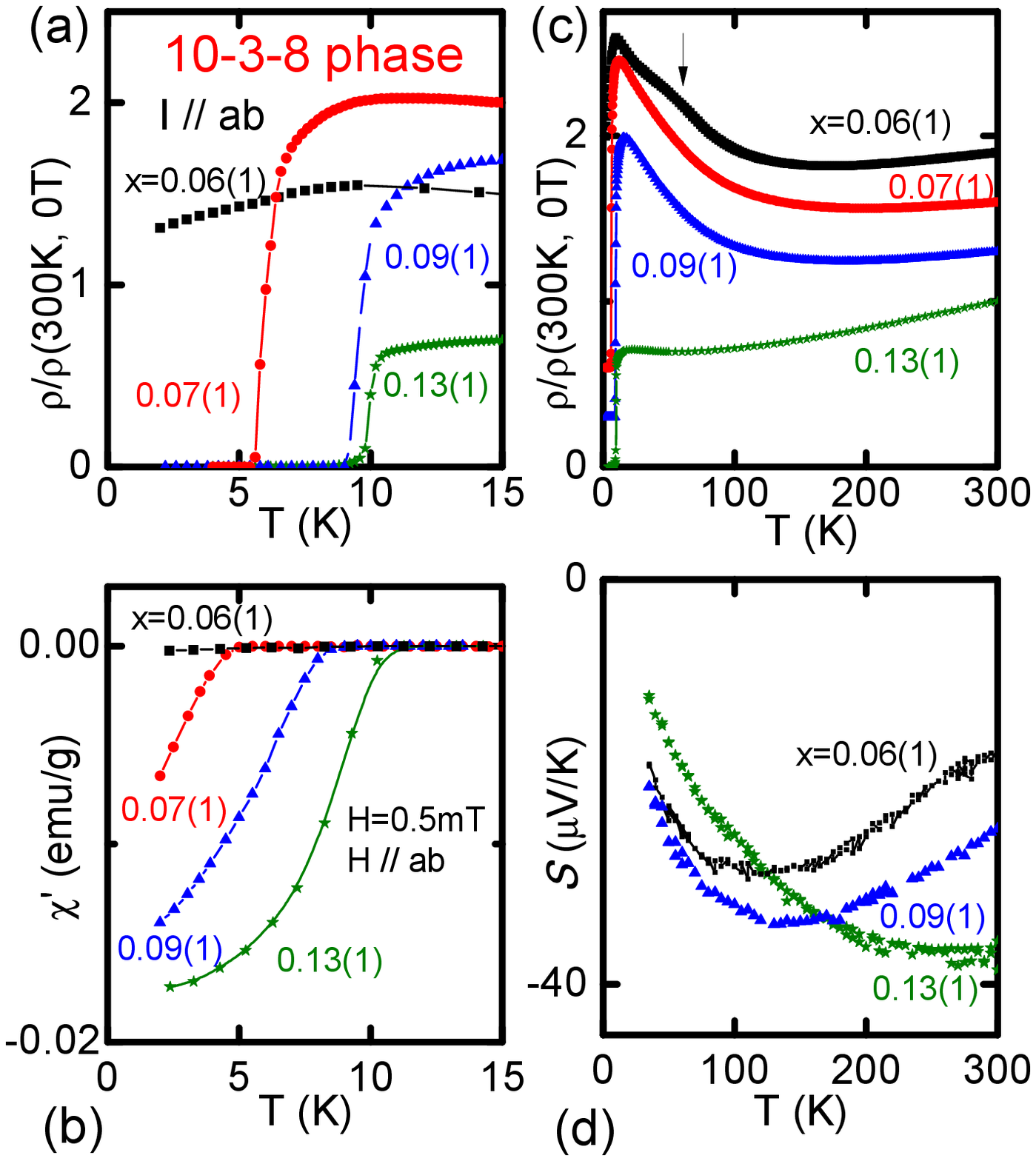,width=3.4in}
\caption{The characterization of the doping-dependent properties for the 10-3-8 phase Ca$_{10}$(Pt$_3$As$_8$)((Fe$_{1-x}$Pt$_x$)$_2$As$_2$)$_5$. (a) The enlarged normalized resistivity,
$\rho(T)/\rho(300~ \rm K)$, near $T_c$ for the 10-3-8 series.
(b) Temperature-dependent AC susceptibility data, $M(T)/H$, taken at 0.5 mT with $H\parallel ab$ in the 10-3-8 phase.
(c) The normalized resistivity, $\rho(T)/\rho(300~ \rm K)$, from 2 to 300 K.
Each subsequent data set is shifted upward by 0.3 for clarity.
(d) The Seebeck coefficient $S(T)$.}
\label{1038}
\eef

We have been successful in tuning the ground state of the 10-3-8 phase from normal to superconducting by controlling the Pt concentration.
The effect of Pt doping in the 10-3-8 phase is summarized in Fig. \ref{1038}.
Because the samples are easily exfoliated, and thus the direct comparison of the resistivities among them are
not suitable\cite{nico, tanatar} (the resistivities at 300 K are all in the 1 m$\Omega$-cm range),
the normalized resistivity, $\rho/\rho(300~ \rm K)$,
is employed in the figure.
The enlarged $\rho/\rho(300 ~\rm K)$ near $T_c$ is shown in Fig. \ref{1038} (a).
Zero resistivity was not observed in $x=0.06(1)$ sample, but does appear at higher Pt concentrations; superconductivity
shows up in the $x=0.07(1)$ sample
with $T_c^{50\%}$ = 5.9 K, and increases to $T_c^{50\%}$ = 9.6 K in the $x=0.09(1)$ sample, $T_c^{50\%}$ = 9.9 K in $x=0.13(1)$ sample.
The nature of the bulk superconductivity
in these samples is confirmed by the AC susceptibility data, presented in Fig. \ref{1038} (b),
which shows large diamagnetic throws with similar magnitudes.
The $T_c$s inferred from both types of measurements are consistent with each other and summarized in Table I.
Figure \ref{1038} (c) shows the evolution of $\rho/\rho(300~ K)$ from 2 to 300 K
with doping (From x=0.06(1) to 0.13(1),
each subsequent data set is shifted upward by 0.3 for clarity).
With decreasing temperature, the resistivity of the $x=0.06(1)$ sample slowly decreases to a minimum around 170 K,
and then monotonically increases, a slope break is observed around 60 K.
This resistivity shape is reminiscent of the one in underdoped
Ba(Fe$_{1-x}$Co$_x$)$_2$As$_2$ \cite{nico} which is associated with the structural
and magnetic phase transitions.
No magnetic anomaly was observed from 2 to 300 K for this compound, however.
For $x=0.07(1)$ and 0.09(1) compounds, upon cooling,
the resistivity decreases slowly first and then increases before dropping to zero; no slope break is observed.
The $x=0.13(1)$ sample shows a quite different resistivity shape, which decreases continuously from 300 to 50 K, followed by
a subtle increase and then a drop to zero.
This shape is reminiscent of the nearly-optimally doped Ba(Fe$_{1-x}$Co$_x$)$_2$As$_2$ \cite{nico}.
The Seebeck coefficient data from 40 to 300 K is presented in Fig. \ref{1038} (d).
$S(T)$ are all negative in measured temperature,
implying the dominant role
of the electron carriers. At 300 K, $S$ decreases from -17 $\mu V/K$ for $x=0.07(1)$ to -37 $\mu V/K$ for $x=0.13(1)$,
indicating that the Pt doping is electron doping.
\bef \psfig{file=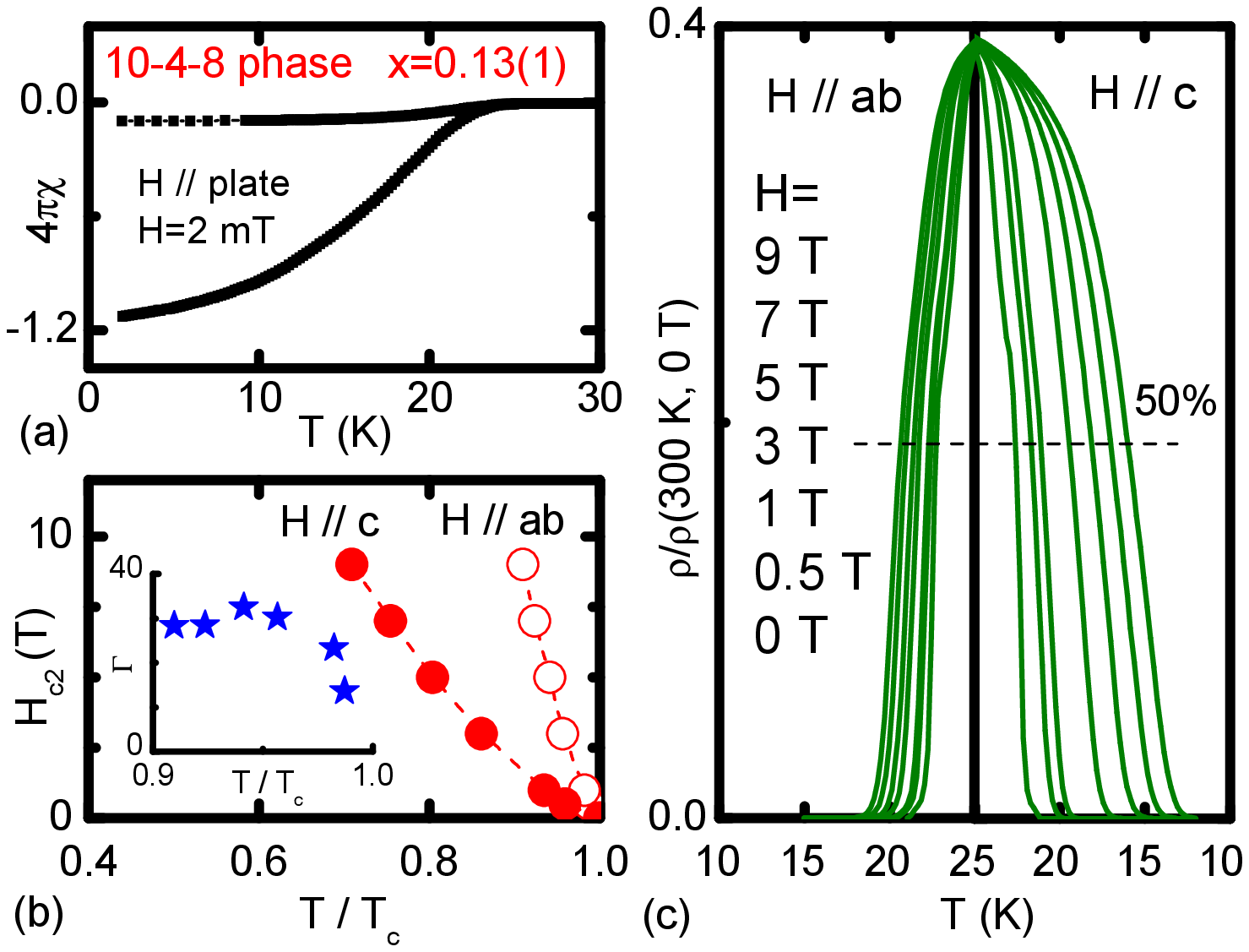,width=3.4in}
\caption{ The characterization of the superconducting properties for the 10-4-8 phase Ca$_{10}$(Pt$_4$As$_8$)((Fe$_{1-x}$Pt$_x$)$_2$As$_2$)$_5$.
(a) The field-cooled / zero-field-cooled DC susceptibility $M(T)/H$ with $H \parallel$ plate.
(b) The normalized resistivity, $\rho/\rho(300 ~\rm K, 0~ \rm T)$ taken at 9, 7, 5, 3,
1, 0.5 and 0 T with $H \parallel ab$ and
$H \parallel c$. The 50\% criterion is shown.
(c) The inferred anisotropic $H_{c2} (T)$ for $H \parallel ab$ and
$H \parallel c$ using the 50\% criterion.
The dashed line is the guide for the eyes. Inset: the anisotropic parameter
$\Gamma (T)=(H_{c2}^{//ab}/H_{c2}^{//c})^2$.}
\label{1048}
\eef

Intergrowth between the PtAs$_2$ and the 10-4-8 phases was detected in a $\sim 0.05 \rm {mm}^2$ area using EDS.
Since we were not able to separate these two phases, neither detailed physical property measurements nor doping studies were
systematically performed. Here we show only the ZFC and FC DC susceptibility data
(Fig. \ref{1048} (a)) and the anisotropic $H_{c2}$ measurement (Fig. \ref{1048} (b) and (c))
for the $x=0.13(1)$ 10-4-8 compound.
The sample used in the $\chi(T)$ measurement
was cut into a plate-like shape and the magnetic field was applied along the plate to minimize the demagnetization effect.
$T_c$ is inferred as 22.5 K.
The estimated shielding fraction is around 120\% (assuming no PtAs$_2$ impurity),
indicating the bulk superconductivity. A thin plate around
 $0.3 \times 0.2 \times 0.005~\rm{mm}^3$ was chosen for the anisotropic $H_{c2}$ measurement.
 The normalized resistivity, $\rho(T)/\rho(300 ~\rm K, 0~ \rm T)$, with field along $ab$ plane and
 $c$ axis is shown in Fig. \ref{1048} (b). At 0 T, $T_c^{50\%}$ is 24.6 K.
 Compared with the 10-3-8 phase, less broadening was observed under the field. The $H_{c2}$
 data was inferred using the 50\% criterion and summarized in Fig. \ref{1048} (c).
 Upon cooling, concave $H_{c2}$ curves were obtained with $H \parallel ab$ and $H \parallel c$.
 Unlike the 10-3-8 phase, this concave shape is criterion-free.
This non-traditional WHH shape is the feature of a two-band superconductor, such as MgB$_2$ \cite{mgb2} and LaFeAsO$_{89}$F$_{0.11}$ \cite{naturehc2}.
The $T_c$ was only suppressed to 0.9$T_{c0}$ with $H \parallel ab$ and 0.7$T_{c0}$  with $H \parallel c$ at 9 T.
Comparing with the $x=0.13(1)$ 10-3-8 phase, less anisotropy was observed.
 The anisotropic parameter $\Gamma$ shown in the inset is 30 at 0.9$T_{c0}$ and 15 near $T_{c0}$ while this value is
25 at 0.9$T_{c0}$ and 100 near $T_{c0}$ in $x=0.13(1)$ 10-3-8 phase.

\section{Discussion}
Although systematic composition dependent physical properties measurements of the 10-4-8 phase are currently not available,
the lower $T_c$ triclinic 10-3-8 phase can be readily compared to the other
Fe pnictides: 1) from 80 to 300 K, a linear temperature dependence of $M(T)/H$ has also been observed in all
the 1111 and 122 Fe pnictide superconductors,
such as LaFeAs(O$_{1-x}$F$_x$) and Ba(Fe$_{1-x}$Co$_x$)$_2$As$_2$ \cite{nico, wang1}.
In the latter case, the linear temperature dependence persists up to 700 K, the highest temperature measured.
It has been suggested that the linear temperature dependence of $M(T)/H$ is related to antiferromagnetic
spin fluctuations \cite{lineartheo}.
2) upon doping, the resistivity changes from a ``semiconducting" appearance in the underdoped region to a ``metallic" appearance near the
optimal doping concentration.
This systematic change is reminiscent of what is seen in other Fe pnictides, where the ``semiconducting" appearance comes from the formation of
a SDW gap \cite{jacs,nico}. However, no magnetic anomalies implying a magnetic phase transition were observed from 2 to 300 K in the 10-3-8 phase.
It is possible that the doping level is still too high, even in the lowest Pt-content samples, and that the SDW transition has already been
fully suppressed.
Further investigation is needed to clarify the ground state of the undoped or very underdoped phase.
3) The anisotropy parameter $\Gamma=m_c^*/m_{ab}^*=(H_{c2}^{//ab}/H_{c2}^{//c})^2$ ranges from 100 near $T_{c0}$ to 25 at 0.9$T_{c0}$.
This is much larger than the 1 to 4 in
(Ba$_{0.55}$K$_{0.45}$)Fe$_2$As$_2$ \cite{mielke} and even larger than the ones in RFeAsO$_{0.8}$F$_{0.2}$ \cite{naturehc2},
and indicates a highly anisotropic 2D nature for the 10-3-8 superconductor.

The $T_c$ in the $x=0.13(1)$ 10-4-8 phase is almost twice the highest $T_c$ we observed in the 10-3-8 phase.
Although we have not yet successfully overdoped the 10-3-8 phase,
the almost linear normal-state resistivity of the $x=0.13(1)$ 10-3-8 compound,
indicates that this compound is very
close to the optimal doping,
and thus represents the nearly maximum $T_c$ obtainable in the triclinic 10-3-8 phase in the Ca-Pt-Fe-As system.
Although we never observed a $T_c$ higher than 25 K in the 10-4-8 phase,
a $T_c$ of 36 K was reported by Kudo et al., for a phase of unreported formula with a crystallographic cell corresponding to our
tetragonal 10-4-8 phase \cite{conference}, suggesting that
$T_c$ can be even higher in the 10-4-8 phase when an optimal doping condition is obtained.

These two chemically and structurally similar compounds provide a particularly interesting platform for studying
superconductivity in the pnictides.
From the chemical point of view, we can straightforwardly model the effective charges in these compounds. The common assignments of
Fe$^{2+}$ and As$^{3-}$ ions in the iron arsenic layer gives the chemical formula [(Fe$_2$As$_2$)$_5$]$^{10-}$.
In the platinum arsenide layer, due to the formation of strong As-As dimers in both compounds, the As in Pt$_n$As$_8$ layer contributes
1 electron to the As-As bond, forming As$_2$$^{4-}$.
For reduced compounds such as skutterudites and
those studied here
the stability of the d$^8$ configuration for Pt rules, and the only oxidation state observed for platinum is +2; this $d^8$
configuration is consistent with the observed Pt-As
coordination polyhedra in the superconductors. This yields an intermediary layer of (Pt$_3$As$_8$)$^{10-}$ for the 10-3-8 phase
and (Pt$_4$As$_8$)$^{8-}$ for the 10-4-8 phase. This results in a major chemical and electronic difference between the two
superconductors: Ca$_{10}$(Pt$_3$As$_8$)(Fe$_2$As$_2$)$_5$ is a
valence satisfied compound through the Zintl concept - the negative charge on the Pt and Fe layers is perfectly balanced by the 20+ charge from the ten electropositive Ca$^{2+}$ ions. In its undoped form the Pt-As layer in the 10-3-8 phase is perfectly charge balanced and therefore will be semiconducting
(i.e., it will not contribute density of states at $E_f$). Ca$_{10}$(Pt$_4$As$_8$)(Fe$_2$As$_2$)$_5$ on the other hand
has one more Pt atom in the Pt-As intermediary layer, thus exceeding its valence satisfaction requirements, indicating that this layer will have states at $E_f$
and is therefore metallic in character.
Thus, given the overall highly similar structures and chemistry,
we argue that the higher $T_c$ in the 10-4-8 phase arises
because the metallic nature of its intermediary layer leads to a stronger interlayer coupling.
Enhanced coupling through intermediary layers has been suggested as the origin of the very high transition temperatures in the highest
T$_c$ cuprates \cite{cup}, but in the iron arsenide superconductors this is the first system where this possibility can be tested.

Structural differences may also have an impact on the superconductivity, but the usual considerations in the arsenide superconductors
do not explain the observations in the present materials.
With a few exceptions, an empirical rule between $T_c$ and $\alpha _{\rm{As-Fe-As}}$, the As-Fe-As bond angle, or $Pn$ height, the distance
between the adjacent Fe and As layer, has been observed the Fe arsenides \cite{angle, johnston, height}. This empirical rule implies that $T_c$ is
enhanced for As-Fe-As bond angles near that of an ideal tetrahedron, 109.47$^\circ $, or for particular values of $P_n \sim 1.38$ \AA\ .
For our compounds, $\alpha _{\rm{As-Fe-As}}$ in the 10-3-8 phase has an average value of 108.99$^\circ $ while $\alpha _{\rm{As-Fe-As}}$ in
the 10-4-8 phase has an average value of 107.52$^\circ $.
Since the 10-4-8 phase has a higher $T_c$ than the 10-3-8 phase, our compounds do not follow the empirical rule of angles. Similarly,
the average $Pn$ is 1.40 \AA\ in the 10-3-8 phase and 1.43 \AA\ in the 10-4-8 phase, which also does not follow the empirical rule of heights.
Thus the structural differences usually credited with governing the $T_c$ in the arsenide superconductors cannot be operating in our phases,
supporting our argument that it is the metallicity and resulting enhanced intralayer coupling in the 10-4-8 phase that determines its higher $T_c$.

Significant differences in the Pt-As interactions between Pt$_3$As$_8$ / Pt$_4$As$_8$ intermediary layers and the neighboring Fe$_2$As$_2$
layers are also present, further strengthening our argument about the importance of the interlayer coupling in determining the $T_c$s.
In the 10-3-8 phase there is only one such interlayer Pt-As interaction per unit cell,
and it appears to be random whether the Pt-As bond points ``up" or ``down". This helps to electronically isolate intermediary layer in
the 10-3-8 phase from the Fe$_2$As$_2$ layers, reinforcing the
perfectly satisfied Zintl valence electron count in this layer, making it more electronically blocking.
In the 10-4-8 compound, on the other hand, the metallic Pt$_4$As$_8$ layer electron count is reinforced by stronger coupling
to the (Fe$_2$As$_2$) layers due to two rather than one bridging Pt-As interactions, and the overall tetragonal symmetry of the phase, which promotes
good orbital overlap between layers. The anisotropy of the properties in the superconducting state,
which show the 10-3-8 phase to be much more anisotropic than the 10-4-8 phase support the chemical picture.
Thus though one can infer that although both compounds should show strong 2D character in their electronic
structures, the 10-4-8 compound will exhibit more and better hybridized electronic states associated with interlayer interactions
and will have an intrinsically metallic intermediary layer, which we propose leads to its much higher T$_c$. This suggests that further
searches for superconducting iron arsenide phases with metallic intermediary layers rather than the commonly found insulating
intermediary layers may be a fruitful path for obtaining higher T$_c$s in the pnictide superconductor family.


\section*{Acknowledgments}
The authors would like to thank E. Climent-Pascual, S. Jia, S. Dutton,
M. Bremholm, M. Fuccillo, M. Ali, J. Krizan, K. Baroudi and H. W. Ji for helpful discussions.
This work was supported by the AFOSR MURI on superconductivity.


\begin{table}
\caption{The crystal structure of the 10-3-8 phase at 100 K. $R1 =  0.0450$ and $wR2 = 0.1102$ for [$F_0 > 4\sigma F_0$] and
$R1 = 0.0902$ for all reflections.
*: the single crystal used here was chosen from the $x=0.07(1)$ batch.}
\begin{ruledtabular}
\begin{tabular}{c c c c c}

  \multicolumn{5}{c}{The 10-3-8 phase $^*$}\\
   \hline
\multicolumn{1}{l}{Crystal system}       & \multicolumn{1}{l}{Triclinic}     &~  & \multicolumn{1}{l}{Sample size}          & \multicolumn{1}{l}{$0.034\times0.063\times 0.067$ mm$^3$}
 \\
\multicolumn{1}{l}{Space group}          & \multicolumn{1}{l}{$P-1$ (\# 2)}  &~  & \multicolumn{1}{l}{Total reflection}       & \multicolumn{1}{l}{2229} \\
\multicolumn{1}{l}{$Z$}                  &\multicolumn{1}{l}{1}              &~  & \multicolumn{1}{l}{Absorption coefficient} & \multicolumn{1}{l}{39.203 / mm} \\
\multicolumn{1}{l}{Unit cell parameters} & \multicolumn{4}{c}{$a$=8.759(4) \AA\ , ~ $b$=8.759(4) \AA\ ,  ~ $c$=10.641(5) \AA\ , $V$=788.1(6) \AA$^3$\ }\\
                                        & \multicolumn{4}{c}{$\alpha $=94.744(5)$^\circ $, ~ $\beta$=104.335(5)$^\circ $,  $\gamma$=90.044(3)$^\circ $ } \\
 \hline
\multicolumn{5}{c}{Atomic position}\\
 \hline
 site &  Wyck & $x/a$         & $y/b$        & $z/c$          \\
  Ca1  &  \emph{2i}   &   0.3655(4)   &  0.1218(4)   &  0.2330(4)     \\
  Ca2  &  \emph{2i}   &     0.7745(4) &  -0.0751(4)  & 0.2352(4)      \\
  Ca3  &  \emph{2i}   &   0.4386(4)   &  0.4798(4)   &  0.7947(4)      \\
  Ca4  &   \emph{2i}  &     0.0284(4) & 0.6810(4)    &   0.7658(4)       \\
  Ca5  &   \emph{2i}  &   0.1688(4)   & 0.7278(4)    &  0.2336(4)        \\
  Fe1  &  \emph{2i}   &      0.1521(7)& 0.5495(8)    &    0.5001(6)      \\
 Fe2   &   \emph{2i}  &      0.2521(7 &  0.2518(7)   &  0.4986(6)      \\
Fe3    &   \emph{2i}  &     0.3494(5) &  -0.0501(6)  &   0.5004(4)    \\
 Fe4   &    \emph{2i} &     0.0498(7) &   -0.1517(7) &  0.4977(5)      \\
 Fe5   &   \emph{2i}  &     0.4516(8) &   0.6481(9)  &  0.5012(8)      \\
  Pt1  &  \emph{2i}  &  0            &   1/2        &0                \\
   Pt2 &  \emph{2i}   &    1/2        & 0            &  0              \\
  Pt3  & \emph{2i}    &  -0.01687(17)  &-0.00552(17)   & -0.05612(16)   \\
  As1  &   \emph{2i}  &   0.11089(19) & 0.03715(19)  &0.36734(18)      \\
As2    &  \emph{2i}   &  0.50913(19)  & -0.16606(19) &   0.36329(18)   \\
As3    & \emph{2i}    &   0.70762(19) & 0.23873(19)  & 0.36320(18)     \\
As4    & \emph{2i}    &  0.68943(19)  & 0.56295(19)  & 0.63737(17)     \\
As5    & \emph{2i}    &   0.09241(19) & 0.36517(19)  & 0.63661(18)     \\
As6    &    \emph{2i} &     0.7344(2) & 0.40095(19)  &  0.00012(19)    \\
As7    &   \emph{2i} &      0.4011(2)& 0.2652(2)    & -0.00063(19)    \\
As8    &    \emph{2i} &      0.2432(2)  & 0.8819(2)    & -0.0001(2)      \\
As9    &    \emph{2i} &      0.1184(2)& 0.2435(2)    & -0.0001(2)      \\
   \hline
\end{tabular}
\end{ruledtabular}
\end{table}

\begin{table}
\caption{The crystal structure of the 10-4-8 phase at 100 K. $R1 =  0.0448$ and $wR2 = 0.1201$ for [$F_0 > 4\sigma F_0$] and
$R1 = 0.0500$ for all reflections
*: the single crystal used here
was grown from a melt containing Ru, with approximately 9\% of Ru and 7\% of Pt on Fe sites.}
\begin{ruledtabular}
\begin{tabular}{c c c c c}

\multicolumn{5}{c}{The 10-4-8 phase$^*$}\\
\hline
\multicolumn{1}{l}{Crystal system}       & \multicolumn{1}{l}{Tetragonal}       &~  & \multicolumn{1}{l}{Sample size}            & \multicolumn{1}{l}{$0.162\times0.109\times 0.082$ mm$^3$} \\
\multicolumn{1}{l}{Space group}          & \multicolumn{1}{l}{$P 4/n$ (\# 85)}  &~  & \multicolumn{1}{l}{Total reflection}       & \multicolumn{1}{l}{823} \\
\multicolumn{1}{l}{$Z$}                  &\multicolumn{1}{l}{1}                 &~  & \multicolumn{1}{l}{Absorption coefficient} & \multicolumn{1}{l}{42.222 / mm} \\
\multicolumn{1}{l}{Unit cell parameters} & \multicolumn{4}{c}{$a$=8.7257(18) \AA\, ~ $b$=8.7257(18) \AA\ , ~ $c$=10.4243(22) \AA\ , $V$=793.7(2) \AA$^3$\ }\\
  ~          & \multicolumn{4}{c}{$\alpha $=90$^\circ $, ~ $\beta$=90$^\circ $,  $\gamma$=90$^\circ $ } \\
 \hline
\multicolumn{5}{c}{Atomic position}\\
 \hline
site  &  Wyck      & $x/a$      & $y/b$               & $z/c$          \\
 Ca1  &  \emph{2c}  &   1/4     &   1/4              & 0.7736(5)      \\
 Ca2 &  \emph{8g}  & 0.3454(2)  &   0.9520(2)        &     0.2417(3)  \\
 Fe1 &   \emph{2b} &   3/4      &  1/4               &  1/2           \\
Fe2  &  \emph{8g}  &   0.457(2) &   0.155(3)         &  0.502(3)      \\
Pt1  & \emph{ 2c}  & 1/4        &   1/4              &    0.06923(9)  \\
Pt2  &  \emph{2a}  &   1/4      &   3/4              &  0             \\
As1  &  \emph{8g}  & 0.54824(10)&  0.34999(11)       &  0.63637(13)   \\
As2  &  \emph{2c}  & 1/4        &  1/4               &  0.3603(2)     \\
As3  &  \emph{8g } & 0.50954(14)&  0.85900(12)       &  0.01687(12)   \\
   \hline
\end{tabular}
\end{ruledtabular}
\end{table}

\end{document}